\documentclass{emulateapj}
\usepackage{graphicx,graphics,amsmath}

\def\bl{Babcock--Leighton}
\def\ftdm{flux transport dynamo model}
\def\Rs{R_{s}}

\renewcommand{\vec}[1]{\mbox{\boldmath $#1$}}

\newcommand{\etasurf}{\eta_{\mathrm{surf}}}
\newcommand{\etaSCZ}{\eta_{\mathrm{SCZ}}}
\newcommand{\etaRZ}{\eta_{\mathrm{RZ}}}
\newcommand{\rBCZ}{r_{\mathrm{BCZ}}}
\newcommand{\rsurf}{r_{\mathrm{surf}}}

\newcommand{\ov}{\overline}

\def\lsim{\lower.5ex\hbox{$\; \buildrel < \over \sim \;$}}
\def\gsim{\lower.5ex\hbox{$\; \buildrel > \over \sim \;$}}
\def\ch{\lower-0.55ex\hbox{--}\kern-0.55em{\lower0.15ex\hbox{$h$}}}
\def\lh{\lower-0.55ex\hbox{--}\kern-0.55em{\lower0.15ex\hbox{$\lambda$}}}

\begin{document}

\title{A dynamo model of magnetic activity in solar-like stars with different
rotational velocities}
\author{Bidya Binay Karak$^{1,2}$, Leonid L. Kitchatinov$^{3,4}$ and Arnab Rai Choudhuri$^1$}
\affil{$^1$Department of Physics, Indian Institute of Science, Bangalore 560012, India\\
$^2$Nordita KTH Royal Institute of Technology and Stockholm University, Roslagstullsbacken 23, SE-106 91 Stockholm, Sweden\\
$^3$Institute for Solar-Terrestrial Physics, Lermontov Str. 126A, Irkutsk 664033, Russia\\
$^4$Pulkovo Astronomical Observatory, St. Petersburg 176140, Russia\\}

\begin{abstract}

We attempt to provide a quantitative theoretical explanation for the observations that Ca II H/K
emission and X-ray emission from solar-like stars increase with decreasing Rossby number (i.e.,\
with faster rotation). Assuming that these emissions are caused by magnetic cycles similar
to the sunspot cycle, we construct flux transport dynamo models of $1M_{\odot}$ stars rotating
with different rotation periods. We first compute the differential rotation and the meridional
circulation inside these stars from a mean-field hydrodynamics model.  Then these are substituted
in our dynamo code to produce periodic solutions.  We find that the dimensionless amplitude $f_m$ of the
toroidal flux through the star increases with decreasing rotation period.  The observational data
can be matched if we assume the emissions to go as the power 3--4 of $f_m$. Assuming that
the Babcock--Leighton mechanism saturates with increasing rotation, we can provide an explanation for
the observed saturation of emission at low Rossby numbers.  The main failure of our model
is that it predicts an increase of magnetic cycle period with increasing rotation rate,
which is the opposite of what is found observationally.  Much of our calculations are based
on the assumption that the magnetic buoyancy makes the magnetic flux tubes
to rise radially from the bottom of the convection zone.  On taking account
of the fact that the Coriolis force diverts the magnetic flux tubes to rise parallel to the
rotation axis in rapidly rotating stars, the results do not change qualitatively.
\end{abstract}

\maketitle
\email{bbkarak@nordita.org}

\section{Introduction}
All late-type stars have convection zones in their outer layers and are expected
to have magnetic activity due to the dynamo action taking place there.
Since the dynamo action crucially depends on rotation (Parker 1955), the
more rapidly rotating stars are likely to have more magnetic activity.
Amongst the important observational signatures
of such stellar magnetic activity are the enhanced Ca II H
and K emission (Noyes et al.\ 1984a; Saar \& Brandenburg 1999) as well as the X-ray emission
(Pallavicini et al.\ 1981; Pizzolato et al.\ 2003; Wright et al.\
2011), there being a good correlation between these two kinds of emission
(Schrijver et al.\ 1992).  Noyes et
al. (1984a) realized that the Rossby number (the ratio of the rotation period to
the convective turnover time) is a particularly convenient parameter to classify
the late-type stars.  In spite of some scatter, Fig.~8 of Noyes et al.\ (1984a)
shows that the Ca II H and K emission has a functional dependence on the Rossby
number---first increasing rapidly with faster rotation and then increasing more
slowly for stars rotating very fast.  Wright et al.\ (2011) present a similar
pattern for the X-ray emission from the late-type stars, as seen in the right
panel of their Fig.~2.  They conclude that there is a power-law relation between
the X-ray emission and the Rossby number for slowly rotating stars (the index
being about $-$2.70), whereas the X-ray emission saturates for rapidly rotating
stars.

Apart from the suggestion that these observational data can be qualitatively explained
by assuming that the dynamo action becomes stronger with rotation and then probably
saturates at sufficiently high rotation, detailed calculations so far have not been done.
During the last few years, models of the solar dynamo have become increasingly
realistic (see, for example, Charbonneau 2010; Choudhuri 2011; and references therein). The aim
of the present paper is to extend our knowledge of the solar dynamo to solar-like
stars rotating at different rates and explore whether the patterns of Ca II H/K and
X-ray emission can be explained quantitatively.

The flux transport dynamo model has emerged in recent years as the
most popular theoretical model of
the Sun's magnetic activity cycle.  In this model, the toroidal
field is produced by the stretching of the poloidal field by differential
rotation at the bottom of the convection zone, where helioseismology has
discovered a concentrated layer of differential rotation known as the tachocline.
The toroidal field produced at the bottom of the convection zone rises to
the solar surface due to magnetic buoyancy.  The decay of tilted sunspot
pairs at the solar surface gives rise to the poloidal field by what is called
the Babcock--Leighton mechanism (Babcock 1961; Leighton 1964) which has received strong
observational support recently (Dasi-Espuig et al. 2010; Kitchatinov \& Olemskoy 2011a).
The meridional
circulation of the Sun plays a very important role in such dynamo models. It
is observed to be poleward at the solar surface and carries the poloidal field
poleward with it. To avoid piling up of matter near the sun's poles, the meridional
circulation has to have an equatorward return flow through the deeper layers
of the convection zone. It is necessary for theoretical models to have such an
equatorward meridional circulation at the bottom of the convection zone to cause
the equatorward advection of the toroidal field generated there, in order to
explain the appearance of sunspots at lower latitudes with the progress of the
solar cycle (Choudhuri et al.\ 1995).

It is expected that solar-type stars rotating at different rates will have
similar flux transport dynamos operating within their convection zones. As
indicated in the last paragraph, we need a detailed knowledge of the differential
rotation and the meridional circulation to model such a flux
transport dynamo. For the Sun, helioseismology has
provided a detailed map of the differential rotation (Schou et al.\ 1998),
which is used in solar dynamo models. Although helioseismology provides some
information about the nature of the meridional circulation in upper layers
of the solar convection zone, we have no direct observational data about the
return flow of meridional circulation through the deeper layers of the
convection zone (but see Zhao et al.\ 2013; Schad et al.\ 2013 for some helioseismic evidence of this).
Since the stable stratification in the radiation zone
does not permit a considerable meridional circulation,
we expect the meridional circulation to remain confined within the
convection zone.  By requiring $\nabla. (\rho {\bf v}) = 0$, one is able to
come up with a reasonable distribution of the meridional circulation which,
when used in solar dynamo models, gives results consistent with observations.
Although some observational information about the differential rotation at the surface
of stars is now available (see, for example, Berdyugina 2005; Collier\,Cameron 2007; Strassmeier 2009),
we need detailed
information about differential rotation throughout the convection zone of a
star in order to construct realistic dynamo models.  Such information about
differential rotation or meridional circulation is not available from observational
data for any stars.  So, in order to construct detailed stellar dynamo models, we
need to calculate the differential rotation and the meridional circulation from theoretical
analysis.

Before the flux transport dynamo model of the solar cycle was proposed
in 1990s (Wang et al.\ 1991; Choudhuri et al.\ 1995; Durney 1995), the importance
of meridional circulation in the dynamo process was not generally recognized
and there were some early efforts of constructing $\alpha \Omega$ models of
stellar dynamos without including the meridional circulation (Belvedere et al.\
1980; Brandenburg et al.\ 1994).  Jouve et al.\ (2010) made the first comprehensive attempt
of constructing flux transport models of stellar dynamos by using differential
rotation and meridional circulation on the basis of 3-D hydrodynamic simulations.
Isik et al. (2011) assume an interface $\alpha \Omega$ dynamo to generate magnetic
fields in stars and use the meridional circulation at the stellar surface to advect
the magnetic flux that has emerged there.

Global meridional flow and differential rotation in convection zones of solar-type stars can be modelled by solving jointly the mean-field equations of motion and heat transport \citep{BMT92,kitch95,kuker01,rempel,hotta11}.
Kitchatinov \& Olemskoy (2011b; hereafter KO11)
calculate differential rotation profiles of main-sequence dwarfs having different
masses and different rotation periods. This model automatically
gives rise to a meridional circulation which is essential for angular
momentum balance. For a solar mass star with solar rotation period, this model
gives a differential rotation profile (see Fig.~1 in KO11) remarkably close to what is found in helioseismology. The model also agrees with measurements of the surface differential rotation in rapidly rotating stars by Barnes et al. (2005).
The accompanying meridional circulation consists of a single cell in the convection
zone with poleward flow near the surface and equatorward flow near the bottom of
the convection zone.  In the present work, we use this model to compute the
differential rotation and the meridional circulation of solar-like stars having
different rotation periods.  Then we give these as inputs in a dynamo model
based on the code {\em Surya} developed in Indian Institute of Science (Nandy
\& Choudhuri 2002; Chatterjee et al.\ 2004; Karak 2010; Karak \&
Choudhuri 2011).

In order to avoid too many complications in this initial exploratory paper, we
restrict ourselves only to stars of mass $1M_{\odot}$.  Since stellar rotation
slows down with age (Skumanich 1972), the sequence of solar mass stars with
different rotation periods can also be viewed as a sequence of stars having
different ages. The differential rotation
and the meridional circulation of such stars having different rotation periods
are first obtained from the model of KO11. As the star
is made to rotate faster, the meridional circulation is found to be confined to
the edges of the convection zone (the poleward flow in a narrow layer near the surface
and the equatorward flow in a narrow layer near the bottom). However, we found that
even such a meridional circulation is able to sustain a flux transport dynamo and
we have been able to construct models of the dynamo operating in $1M_{\odot}$ stars
having different rotation periods by putting the appropriate differential rotation
and the appropriate meridional circulation in the dynamo code. The next challenging question
is to connect the results of the dynamo calculation with the observed emission
in Ca II H/K and X-ray.

In order to limit the growth of the magnetic field generated by the dynamo, it is
necessary to include the nonlinear feedback of the growing magnetic field on the
dynamo.  The simplest way of doing this is to include a quenching in the $\alpha$
parameter describing the generation of the poloidal field.  If the quenching is
of such a nature that the dynamo action gets quenched when the toroidal field is
stronger than $B_0$, then the maximum value of the toroidal field hovers around
$B_0$ and the total toroidal flux in the convection zone at an instant would be
$f B_0 \Rs^2$, where $f$ is usually much smaller than 1.  As the toroidal
field changes sign, $f$ is expected to vary in a periodic fashion going through positive
and negative values.  Let $f_m$ be the amplitude of $f$, implying that the maximum
toroidal flux in the convection zone is $f_m B_0 \Rs^2$. In our dynamo
simulations of $1M_{\odot}$ stars with different rotation periods, we find that
$f_m$ increases with decreasing rotation periods.  In other words, stars rotating
faster produce more magnetic flux. The emissions in Ca II H/K or X-ray, which depend
on the overall magnetic activities of the stars, are expected to increase with
increasing $f_m$. Since the emissions (especially the X-ray emission)
often arise from magnetic reconnection involving
the interaction of one magnetic flux system with another, one may naively expect
that the emissions may go as $f_m^2$. One of the aims of the present study is to
check whether the observed dependence of the emissions on the Rossby number can
be explained on the basis of such assumptions.

One assumption used in most of the dynamo models developed so far is that
magnetic buoyancy makes the toroidal field at the bottom of the convection zone to
rise radially through the convection zone to the solar surface. However, in the case of
stars rotating sufficiently fast, the Coriolis force makes flux tubes rise parallel
to the rotation axis rather than radially (Choudhuri \& Gilman 1987; Choudhuri 1989;
D'Silva \& Choudhuri 1993; Fan et al.\ 1993; Weber et al. 2011). This happens when the
rotation period is shorter than the dynamical time scale of the flux tube rise.
If the dynamical time scale of the flux tube rise is comparable to the turnover
time of convection (which is the case for the Sun), then we would expect the
Coriolis force to be dominant and make flux tubes rise parallel to the rotation
axis when the Rossby number is less than 1.  It is intriguing that the nature of
dependence of the emission on Rossby number changes around such a value, as seen
in Fig.~8 of Noyes et al.\ (1984a) and Fig.~2 of Wright et al.\ (2011). The usual
explanation given for such saturation is that the dynamo saturates when the
rotation is fast.  One important question is whether a change in the nature of
the dynamo due to the change in the nature of magnetic buoyancy could also be
behind this.  Doppler imaging shows that some fast rotating stars have
polar spots (Vogt \& Penrod 1983; Strassmeier et al.\ 1991), which are believed to
be caused by the Coriolis force diverting the rising flux tubes to high latitudes
(Sch\"ussler \& Solanki 1992). But, to the best of our knowledge, no previous
calculation has been done to explore how the nature of the dynamo changes
when the magnetic flux rises parallel to the rotation axis rather than radially.
We present some such calculations and show that the dynamo becomes less efficient
and generates less magnetic flux when the toroidal field is assumed to rise
parallel to the rotation axis.

One important question connected with the theory of stellar dynamos is to obtain
periods of magnetic cycles.  Thanks to the program of the Mount Wilson Observatory
for monitoring Ca H/K emission from several solar-like stars for many years, the activity cycles
of many solar-like stars have been discovered (Wilson 1978; Baliunas et al.\ 1995).
There is evidence that stars with longer rotation periods have longer activity
cycle periods.  This was first reported by Noyes et al. (1984b), who pointed
out that this can be easily explained for a linear $\alpha \Omega$ dynamo on
the basis of some scaling arguments (see also Brandenburg et al.\ 1998).
However, in flux transport dynamos, the cycle period is determined essentially
by the time scale of the meridional circulation.  Ironically, flux transport
stellar dynamos have difficulty in explaining the observed increase of cycle
period with the increase of rotation period (or with the decrease of rotation rate).
This was pointed out by Jouve et al.\ (2010, see the second panel of their Figure~4),
who discussed various ways of getting around this difficulty. We also encounter
this difficulty in our calculations. However, in the present paper, we do not
discuss possible mechanisms of solving this difficulty.

The mathematical formulation of the flux transport dynamo model is described in
the next Section. In \S3 we present the results obtained by assuming that the flux
tubes rise radially through the convection zone due to magnetic buoyancy. Then we
discuss in \S4 how our results get modified when we allow the Coriolis force to
make the flux tubes rise parallel to the rotation axis.  Our conclusions are
summarized in \S5.

\section{Flux transport dynamo model}

We assume the magnetic field to be axisymmetric and write it in  the following form:
\begin{equation}
{\bf B} = \nabla \times [ A(r, \theta) {\bf e}_{\phi}] + B (r, \theta) {\bf e}_{\phi}
\end{equation}
where ${\bf B_p} = \nabla \times [ A(r, \theta) {\bf e}_{\phi}]$ is the poloidal component of the magnetic field and
$B(r, \theta)$ is the toroidal component.
Then, in the \ftdm, we solve the following equations to study the evolution of the magnetic fields:
\begin{equation}
\frac{\partial A}{\partial t} + \frac{1}{s}({\bf v_p}.\nabla)(s A)
= \eta \left( \nabla^2 - \frac{1}{s^2} \right) A + S(r, \theta; B),
\end{equation}
\begin{eqnarray}
\frac{\partial B}{\partial t}
+ \frac{1}{r} \left[ \frac{\partial}{\partial r}
(r v_r B) + \frac{\partial}{\partial \theta}(v_{\theta} B) \right]
= \eta \left( \nabla^2 - \frac{1}{s^2} \right) B \nonumber \\
+ s({\bf B}_p.{\bf \nabla})\Omega + \frac{1}{r}\frac{d\eta}{dr}\frac{\partial{(rB)}}{\partial{r}},~~~~~~~~~~~~~~~~~~~~~~~~~~~~~
\end{eqnarray}\\
where $s = r \sin \theta$.

Here ${\bf v_p} = v_r {\bf \hat r} + v_{\theta} {\bf \hat \theta}$ is the meridional circulation
velocity, whereas $\Omega$ is the angular velocity, both ${\bf v_p}$ and $\Omega$ being functions
of $r$ and $\theta$. The coefficient $\eta$ is the turbulent magnetic diffusivity. The Babcock--Leighton
mechanism for generating the poloidal field is encaptured through the source term $S(r, \theta; B)$
in Equation~(2).  We shall discuss below how we specify $\eta$, $S(r, \theta; B)$, $\Omega$ and ${\bf v_p}$.
Once these parameters are given, we have to solve Equations~(2) and (3) within a hemisphere of the convection zone
of the star. Since the bottom of the convection zone for solar-like stars is at $0.7 \Rs$, we carry on
the numerical integration of Equations~(2) and (3) within the range $0.55 \Rs < r < \Rs$ and $0 <\theta <\pi/2$,
the bottom of the integration region being a little bit below the bottom of the convection zone.
We now come to the boundary conditions which we use.
The boundary conditions at the pole are $A = 0, B = 0 $, and at the equator are
$\partial A / \partial r = 0, B = 0 $ which force a dipolar symmetry,
whereas at the lower boundary we take $A = 0, B  = 0$.
Above the upper boundary, we assume the magnetic field to be a potential field. The
upper boundary condition we use is $B = 0$ and $A$ smoothly matches this potential field across the boundary
(see Dikpati \& Choudhuri 1995 and Chatterjee et al. 2004 for details). We use the code {\em Surya}
developed at the Indian Institute of Science to solve Equations~(2) and (3) with these boundary conditions
(Nandy \& Choudhuri 2002; Chatterjee et al.\ 2004; Karak 2010; Karak \&
Choudhuri 2011).
We have used spatial resolution of $256\times256$ for all the calculations. The code has
 been run for a sufficiently long time compared to the diffusion time so that the results are
 disregarded for the initial conditions.

\begin{figure}[!h]
    \centering
    \includegraphics[width=0.35\textwidth]{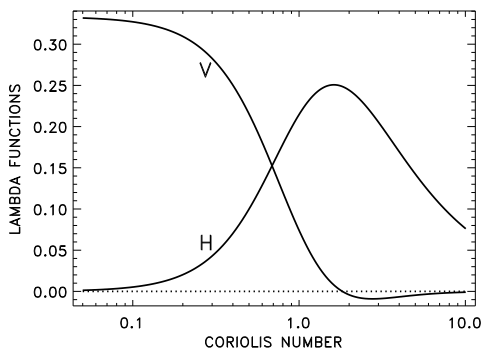}
    \caption{Normalized fluxes $V$ and $H$ of angular momentum
        of the Equation (\ref{Lambda}) as functions of the Coriolis number.}
    \label{VH}
\end{figure}

We now come to the specification of the various parameters. We use the angular velocity profile $\Omega (r, \theta)$ and the meridional circulation ${\bf v_p}$ computed from the mean-field hydrodynamics model of KO11.
The numerical model solves the steady motion equation
\begin{equation}
    \left({\vec V}\cdot{\vec\nabla}\right){\vec V}
    + \frac{1}{\rho}\vec{\nabla}P -\vec{g}
    = \frac{1}{\rho}\vec{\nabla}\cdot\vec{R} ,
    \label{mot}
\end{equation}
where the mean velocity  $\vec V$ combines the axisymmetric rotational and meridional flows,
$\vec{V} = \vec{e}_\phi r\sin\theta\Omega + {\bf v_p}$ and $\vec g$ is gravity. Convective velocities $\vec u$ contribute the motion equation (\ref{mot}) via the Reynolds stress
\begin{equation}
    R_{ij} = -\rho Q_{ij},\ \ Q_{ij} = \langle u_i u_j\rangle ,
    \label{RS}
\end{equation}
where $Q_{ij}$ is the correlation tensor of fluctuating velocities. Differential rotation results in the model from the angular momentum transport by convection and meridional flow. Convective fluxes of angular momentum are proportional to the correlation, $Q_{\phi r}$ and $Q_{\phi\theta}$, of azimuthal and meridional velocities. Apart from the contribution of the eddy viscosities, $Q^\nu_{ij}$, the correlation $Q_{ij} = Q^\nu_{ij} + Q^\Lambda_{ij}$ contains a non-viscous part, $Q^\Lambda_{ij}$, the presence of which in rotating fluids is named the $\Lambda$-effect \citep{R89}.
It is specified after the quasi-linear theory of turbulent transport (cf. \citeauthor{Rea13} \citeyear{Rea13}),
\begin{eqnarray}
    Q^\Lambda_{r\phi} &=& \frac{\tau\ell^4 g}{15H_\rho c_\mathrm{p}}
    \frac{\partial S}{\partial r}\ \Omega\sin\theta
    \left(V(\Omega^*) + H(\Omega^*)\cos^2\theta\right) ,
    \nonumber \\
    Q^\Lambda_{\theta\phi} &=& -\frac{\tau\ell^4 g}{15H_\rho c_\mathrm{p}}
    \frac{\partial S}{\partial r}\ \Omega\sin^2\theta \cos\theta H(\Omega^*),
    \label{Lambda}
\end{eqnarray}
where $\tau$ is the correlation time of convective turbulence, $\ell$ is the correlation length, $H_\rho$ is the density scale height, $c_\mathrm{p}$ is the specific heat capacity at constant pressure, $S$ is the specific entropy, and $V$ and $H$ are dimensionless functions of the Coriolis number $\Omega^* = 2\tau\Omega$ (Figure \ref{VH}).
The entropy gradient appears in the Equation (\ref{Lambda}) because the background turbulence intensity is expressed in terms of the entropy gradient. 
The viscous  part of the correlation reads,
\begin{equation}
    Q^\nu_{ij} = -{\cal N}_{ijkl}\frac{\partial V_k}{\partial r_l} .
    \label{visc}
\end{equation}
The eddy transport coefficients ${\cal N}$ are also expressed in terms of the entropy gradient which gradient in turn is controlled by the entropy equation which is solved together with the motion equation (\ref{mot}). This is done to reduce arbitrariness in specifying the model parameters. There only tunable parameter is $C_\chi$ of the eddy thermal diffusivity,
\begin{equation}
    \chi_{ij} = -\frac{\tau\ell^2 g}{12c_\mathrm{p}}\frac{\partial S}{\partial r}
    \left( \phi (\Omega^*)\delta_{ij} + C_\chi \phi_\| (\Omega^*)
    \hat{\Omega}_i\hat{\Omega}_j\right) .
    \label{C_chi}
\end{equation}
With this equation, the thermal diffusivities differ between the directions along and normal to the rotation axis, i.e., the diffusivity is anisotropic. The anisotropy is induced by rotation. In the limit of slow rotation, $\Omega^* \rightarrow 0$, the second term in the brackets of Equation (\ref{C_chi}) vanishes, $\phi_\| (0) = 0$, and the diffusivity becomes isotropic. Finite anisotropy in a rotating star, however, is very important for its differential rotation structure. The anisotropy results in a differential temperature: a slight increase in mean temperature with latitude. The differential temperature is essential for deviation from the Taylor--Proudman state of cylinder-shaped rotation \citep{Rea13}.

The value of $C_\chi = 1.5$ in Equation (\ref{C_chi}) gives close agreement with helioseismology. The $C_\chi$-parameter was, therefore, fixed to this value and not varied in stellar applications. In order to compute the differential rotation, the model needs the structure of a star to be specified. We use the {\sl EZ} code of stellar evolution by Paxton (2004) to specify the structure of a $1M_\odot$ star as a function of age. Then, the gyrochronology relation by Barnes (2007) was used to identify the rotation rate for a star of given age. This provides a sequence of models for differential rotation and meridional flow in the Sun of different ages. All details about this modeling of stellar differential rotation can be found in KO11 and Kitchatinov \& Olemskoy (2012a). The computations give solar-type rotation with the equator rotating faster than poles. Only with unrealistically slow rotation (with, say, $P_\mathrm{rot} = 60$~days) and artificially reduced anisotropy of heat transport can the anti-solar rotation be found (Kitchatinov \& Olemskoy 2012a). A similar conclusion has been reached 
by Gastine et al.\ (2014) and Karak et al.\ (2014a).

We point out that the model of KO11 does not include overshooting and provides the differential rotation
only above $0.72 \Rs$.  We assume the core of the star below $0.7 \Rs$ to rotate with the constant angular velocity
$\Omega_{\rm core}$ corresponding to the rotation period used for the particular case.  To produce a smooth fit between
$0.7 \Rs$ and $0.72 \Rs$, we use the following procedure. Let $\Omega_{\rm model} (r, \theta)$ be equal to $\Omega$ given by the
model of KO11 above $0.72 \Rs$, whereas below $0.72 \Rs$ we assume $\Omega_{\rm model}(r, \theta)$ to be
equal to the value of $\Omega$ at ($r = 0.72 \Rs, \theta$) from the model. We now take the angular velocity to be given
by the following expression
\begin{eqnarray}
\Omega (r, \theta) = \Omega_{\rm model} (r, \theta) + \frac{1}{2}[\Omega_{\rm core} - \Omega_{\rm model} (r = 0.72 \Rs, \theta)] \nonumber \\
\left[ 1 - {\rm erf} \left( \frac{r - 0.71 \Rs}{0.01 \Rs} \right) \right]~~~~~~
\end{eqnarray}
This expression of angular velocity implies a strong differential rotation between $0.7 \Rs$ and $0.72 \Rs$, which
is our tachocline.

\begin{figure}[!h]
\centering
\includegraphics[width=0.58\textwidth]{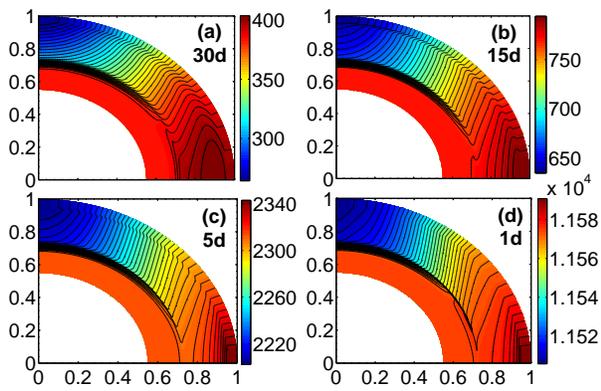}
\caption{The distribution of angular velocity in the poloidal planes of stars with
rotation periods 30, 15, 5 and 1 days. The rotational frequencies in nHz are indicated
by the different colors. Note that the lower boundary at about $0.72\Rs$ of the KO11
model is smoothed using Equation (9) to form a tachocline-like shear layer.}
\label{omega}
\end{figure}

\begin{figure}
\includegraphics[width=1.45in]{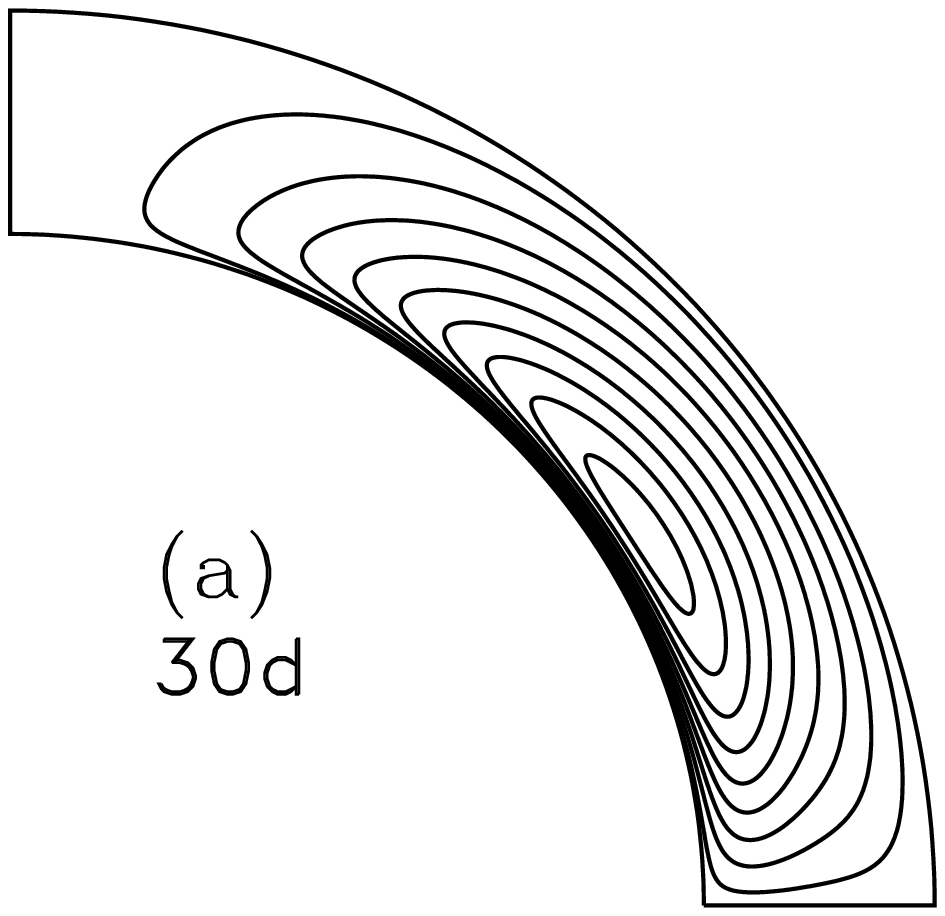}
\includegraphics[width=1.45in]{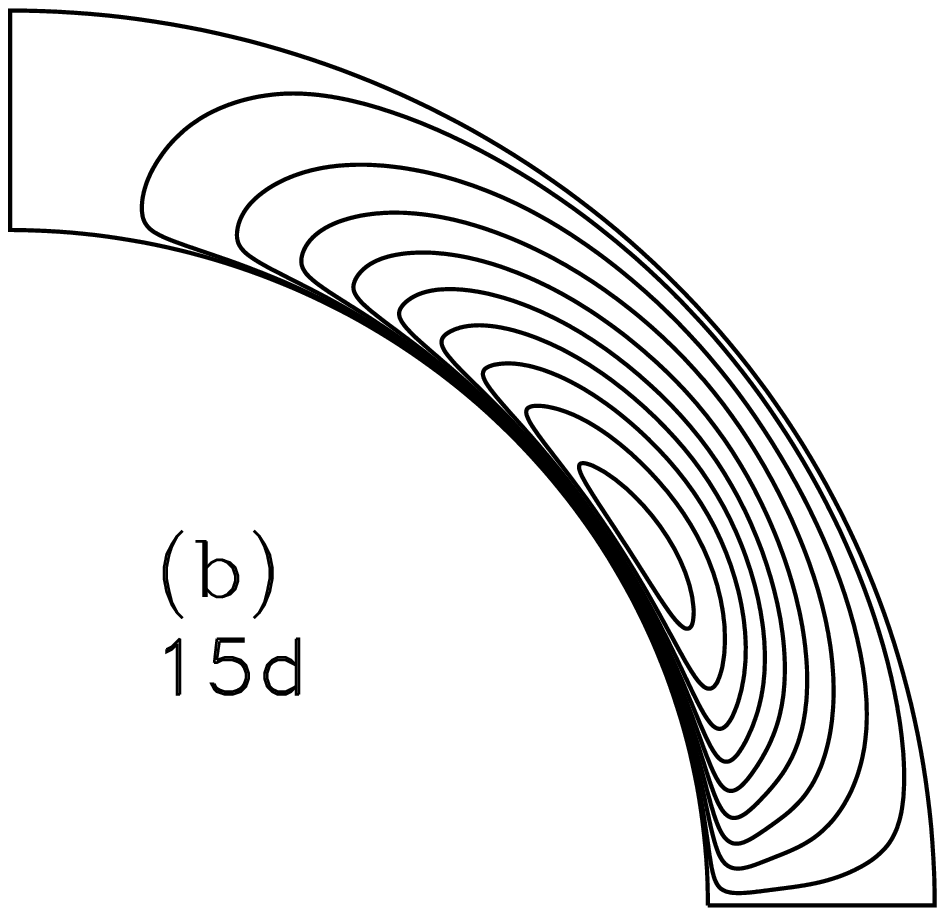}
\includegraphics[width=1.45in]{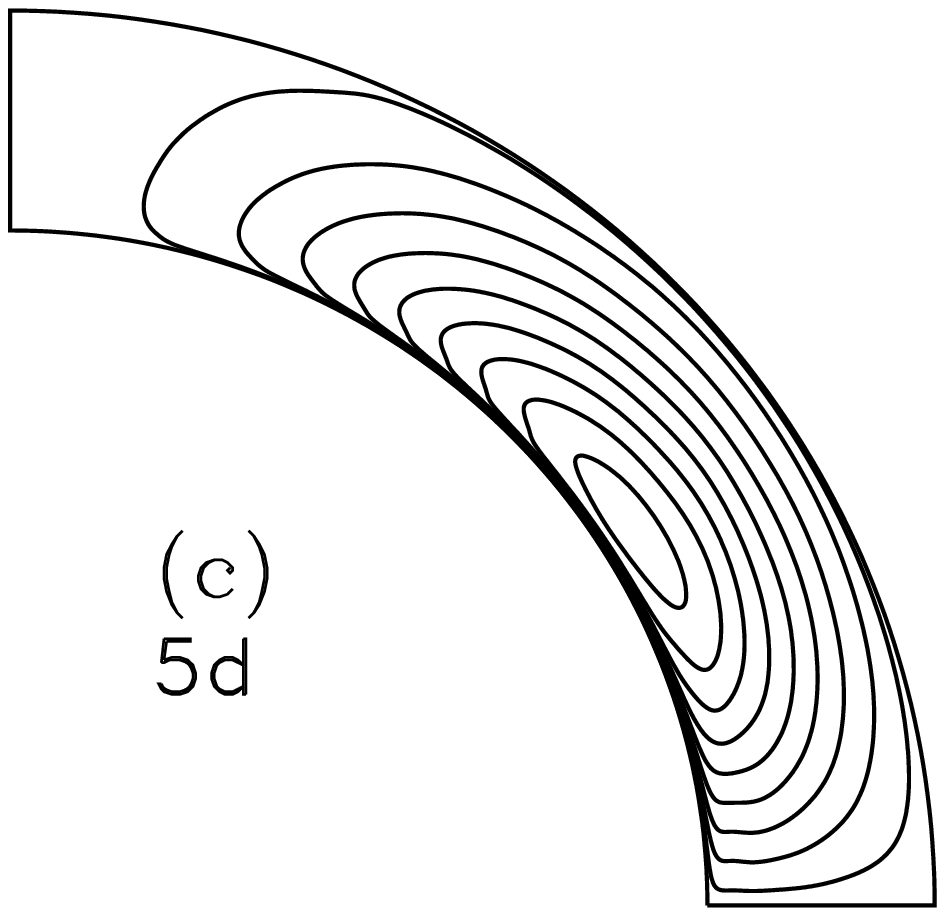}
\includegraphics[width=1.45in]{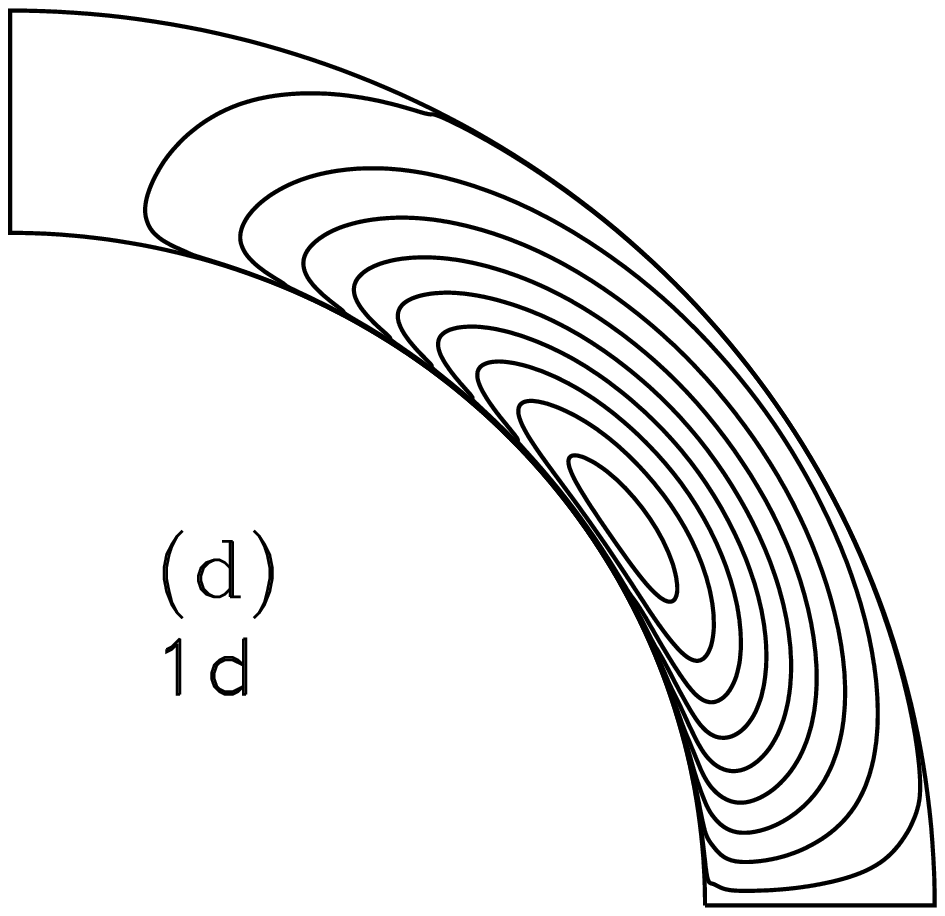}
\caption{Streamlines of meridional circulation of stars with
rotation periods 30, 15, 5 and 1 days.}
\label{fig:stream}
\end{figure}

\begin{figure}[!h]
\centering
\includegraphics[width=0.50\textwidth]{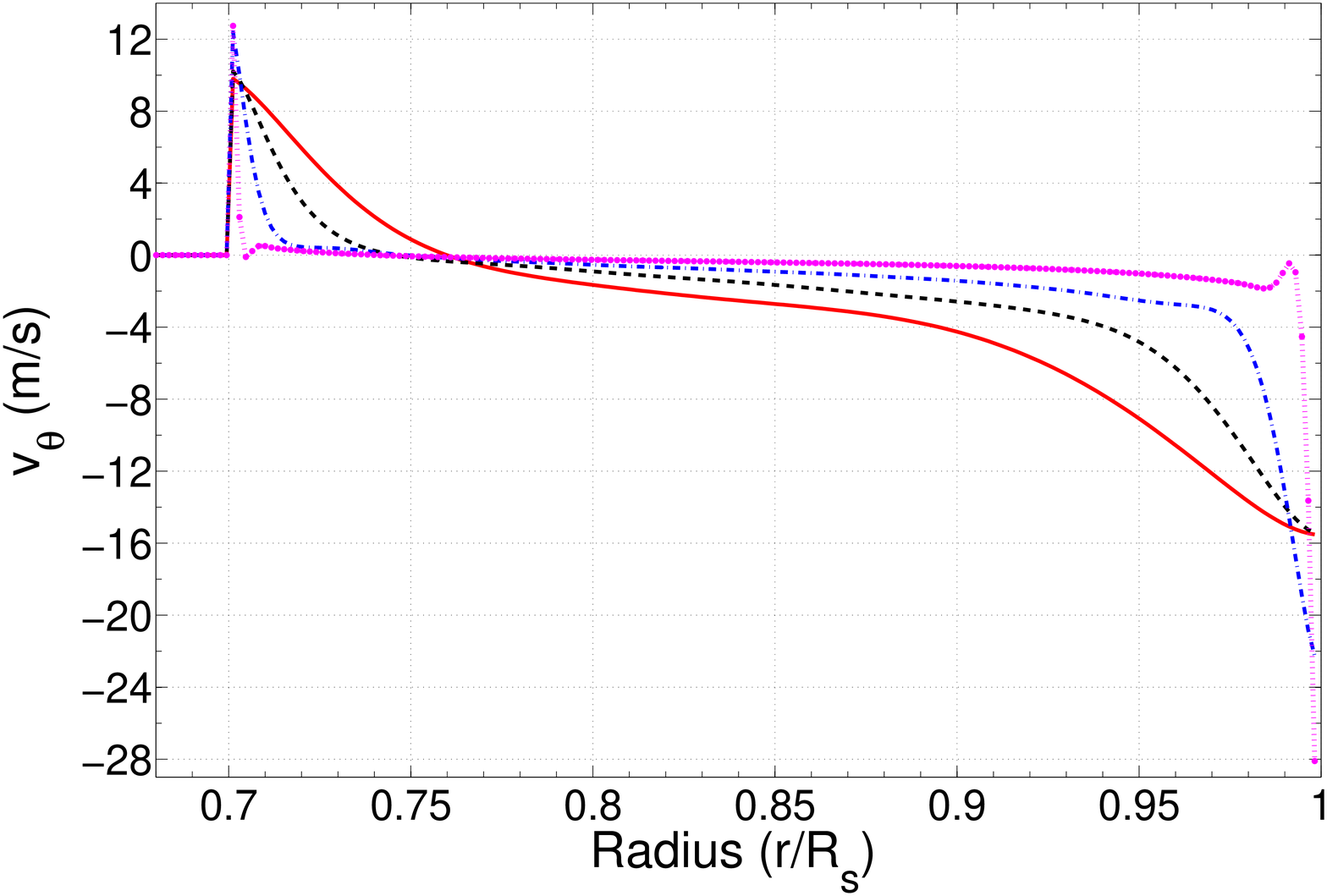}
\caption{The component $v_{\theta}$ (in m~s$^{-1}$) of meridional circulation
at $45^0$ latitude of four different stars. Solid (red),
dashed (black), dash-dotted (blue) and dot-pointed (magenta) lines correspond to
stars with rotation periods 30, 15, 5 and 1 days respectively.}
\label{vtheta}
\end{figure}

In this
paper, we carry on calculations for $1M_{\odot}$ stars having rotation periods of 1, 2, 3, 4, 5, 7, 10, 15,
20, 25.38 (solar value) and 30 days.
Figure~\ref{omega} shows the angular velocity distributions of stars with rotation periods of
1, 5, 15 and 30 days.  It is clear that the angular velocity tends to be constant
over cylinders when the rotation is fast, whereas there is a tendency of it being
constant over cones for rotation periods comparable to the solar rotation period and longer.
For all the cases we computed, the meridional circulation always
consists of a single cell with poleward flow near the surface and equatorward flow at
the bottom of the convection zone. For faster rotations (i.e.\ for shorter rotation
periods), the meridional circulation tends to be confined to the peripheries of the
convection zone.  In Figure~\ref{fig:stream}, we show the streamlines of meridional circulation, 
whereas in Figure~\ref{vtheta}, we show $v_{\theta}$ as function of $r$ at the latitude
$45^{\circ}$ for rotation periods of 1, 5, 15 and 30 days. We point out again that
the model of KO11 does not include overshooting in the tachocline.
As a result, the meridional circulation abruptly falls to zero at the bottom of the
convection zone.

The only remaining parameters to be specified are the turbulent magnetic diffusivity
$\eta$ and the source term $S(r, \theta; B)$.  We specify them in such a way that the
code {\em Surya} gives stable periodic solutions for solar-like stars over the wide range of rotational
velocities we are considering.  Thus we have specified them in a way somewhat different from
what we have done in some of our earlier calculations (Nandy \& Choudhuri 2002; Chatterjee
et al.\ 2004; Karak \& Choudhuri 2011, 2012, 2013). With our earlier specifications
of $\eta$ and  $S(r, \theta; B)$, we had been able to reproduce various aspects of the solar
cycle extremely well.  However, we now find that these earlier specifications do not give stable periodic
solutions when we use the angular velocity profile and the meridional circulation appropriate
for very short rotation periods.  So we use somewhat different specifications of turbulent
diffusivity and magnetic buoyancy, which are very similar to what is done in other
flux transport dynamo models (e.g., Mu\~noz-Jaramillo et al.\ 2009; Hotta \& Yokoyama 2010). When
we use these specifications for the case of solar rotation period, we find the activity cycle period
to be around 6.5 yr instead of 11 yr (see Choudhuri et al.\ 2005 for a discussion)
and the butterfly diagram also does not look very solar-like.
However, here we are interested more in finding out how the behavior of the dynamo changes
with different rotational velocities, rather than matching the observational data for only
one value of rotational velocity (the solar value) for which we have detailed observational data.
So we have used a model in which we can hold the other things invariant while changing the
angular velocity profile and the meridional circulation for different rotation periods. We
hope that the results obtained with this model at least qualitatively captures the changing
behavior of the dynamo with different rotation periods.

\begin{figure}[!h]
\centering
\includegraphics[width=0.35\textwidth]{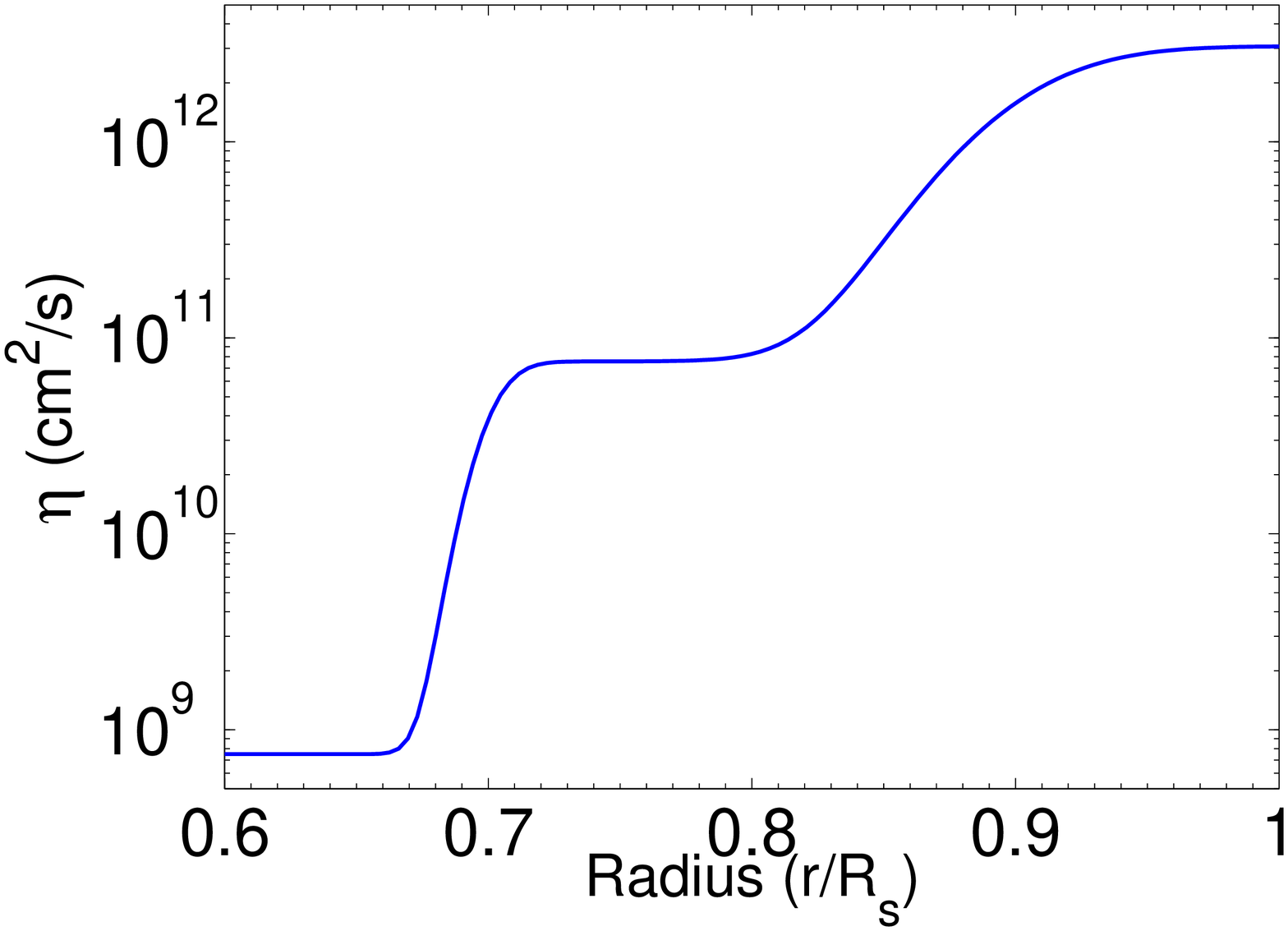}
\caption{Variation of turbulent diffusivity $\eta$ with stellar radius used in our dynamo model.}
\label{eta_prof}
\end{figure}

We take the turbulent magnetic diffusivity to be a function of $r$ alone, having the following form:
\begin{eqnarray}
\eta(r) = \etaRZ + \frac{\etaSCZ}{2}\left[1 + \mathrm{erf} \left(2\frac{r - \rBCZ}
{d_t}\right) \right]\nonumber \\
+\frac{\etasurf}{2}\left[1 + \mathrm{erf} \left(\frac{r - \rsurf}
{d_2}\right) \right]
\label{eq:etap}
\end{eqnarray}\\
with $\rBCZ=0.7 \Rs$, $d_t=0.03 \Rs$, $d_2=0.05 \Rs$, $\rsurf = 0.95 \Rs$,
$\etaRZ = 5 \times 10^8$ cm$^2$~s$^{-1}$, $\etaSCZ = 5 \times 10^{10}$ cm$^2$~s$^{-1}$, and
$\etasurf = 2\times10^{12}$ cm$^2$ s$^{-1}$.
The radial dependence of the turbulent diffusivity is shown in Figure~\ref{eta_prof}. Note that the
diffusivity is weaker in the lower half of the convection zone compared to its value
in the upper layers and it falls to a very low value below the convection zone.
Unlike the thermal diffusivity and viscosity used in the hydrodynamics model of KO11, our turbulent magnetic diffusivity is not derived from any theory but prescribed similar to the diffusivity profile used by Mu\~noz-Jaramillo et al.\ (2009) and Hotta \& Yokoyama (2010).
In our all the calculations we use the same diffusivity.

The source term $S(r, \theta; B)$ captures the Babcock--Leighton mechanism of
the generation of the poloidal field near the stellar surface from the decay of tilted bipolar starspots.
We use the following form for this term:
\begin{equation}
 S(r, \theta; B) = \frac{\alpha(r,\theta)}{1+(\ov B(r_t,\theta)/B_0)^2} \ov B(r_t,\theta),
\label{source}
\end{equation}
where $\ov B(r_t,\theta)$ is the value of the toroidal field at latitude $\theta$
radially averaged over the tachocline from $r = 0.685 \Rs$ to $r=0.715 \Rs$. We take
\begin{eqnarray}
\alpha(r,\theta)=\frac{\alpha_0}{4}\left[1+\mathrm{erf}\left(\frac{r-r_4}{d_4}\right)\right]\left[1-\mathrm{erf}\left(\frac{r-r_5}{d_5}\right)\right]\nonumber\\
\times \sin\theta\cos\theta~~~~~~~~~~~~
\label{alpha}
\end{eqnarray}
with $r_4=0.95 \Rs$, $r_5= \Rs$, $d_4=0.05 \Rs$, $d_5=0.01 \Rs$.
These parameters ensure that $\alpha(r,\theta)$ is non-zero only in a thin
layer near the surface, making the \bl\ mechanism operative only near the stellar surface.
Following many previous authors, we carry on a non-local treatment of magnetic buoyancy
by making the \bl\ mechanism operate on the magnetic field $\ov B(r_t,\theta)$ in the
tachocline. We have also included $\alpha$-quenching.  While the $\alpha$-quenching is easier
to interpret for the traditional $\alpha$-effect based on helical turbulence (Parker 1955; Steenbeck
et al.\ 1966), we expect such quenching to be present even in the \bl\ mechanism,
since a stronger magnetic field reduces the relative importance of the Coriolis force compared to
the magnetic buoyancy, thereby reducing the tilt of the emerging starspot pair (D'Silva \& Choudhuri 1993; Weber et al.\ 2011).
The factor $B_0$ appearing in the quenching is the only nonlinearity in our model. Since the dynamo
action is suppressed when the toroidal field exceeds this value, we expect the maximum value of
the toroidal field in the tachocline not to exceed $B_0$ substantially.
Essentially this $B_0$ should vary with different stars. However to make the calculation simple
and to interpret the results clearly, we take a fixed value of $B_0$ for all the cases.
The factor $\alpha_0$ in (12) determines the strength of the \bl\ process. For shorter rotation
periods, the Coriolis force is stronger, making the \bl\ process also stronger by making
the tilts of bipolar starspots larger.  Taking the \bl\ process to be inversely proportional
to the rotation period $T$, we write
$$\alpha_0 = \alpha_{0,s} \frac{T_s}{T}, \eqno(13)$$
where $T_s$ is the solar rotation period and $\alpha_{0,s}$ is the value of $\alpha_0$
for the solar case, which we take  $\alpha_{0,s} = 1.6$ cm~s$^{-1}$. We shall see in
\S3 that the magnetic activity keeps on rising for shorter rotation periods when we use
Equation~(13), instead of being saturated as seen in observational data. One possible reason behind
the saturation seen in the observational data for short rotation periods is that the dynamo
action gets saturated for very fast rotations.  This can be phenomenologically included
by replacing Equation~(13) by
$$ \alpha_{0}= \frac{\alpha_{0,s}}{\beta} \left[1 - \exp \left(-\frac{\beta T_s}{T} \right) \right]. \eqno(14)$$
When we use such an expression, the \bl\ mechanism saturates when $T \ll \beta T_s$ and we
get back Equation~(13) when $T \gg \beta T_s$.

Finally, it should be noted that magnetic buoyancy is included in Equation (11) on the assumption that
flux tubes rise radially.  As we pointed out, the strong Coriolis force may make flux tubes
rise parallel to the rotation axis when the rotation period is sufficiently short. We shall consider
this possibility in \S4, where we shall have to modify (11).

\section{Results for radial rise of magnetic flux}

We now present the results obtained by using the source term of the form (11), which implies
that magnetic flux rises radially due to magnetic buoyancy.
We first perform some simulations with a fixed \bl\ $\alpha$ and other parameters in the model.
We find that the dynamo becomes weaker with the increase of rotation. The reason behind it is that the
meridional circulation becomes weaker with rotation rate (see Figure 4). In flux transport dynamo model the
weak meridional circulation makes the dynamo weaker due to turbulent diffusion. This effect is
stronger than the increase of the shear with rotation.
In other words, we can say that the $\Omega$-effect from our differential
rotation does not have much role in increasing the magnetic flux.
Therefore, we discuss results in which
the \bl\ mechanism is assumed to vary with rotation period as Equation (13) without a saturation for rapid rotations as
included in (14).  Toward the end of this section, we shall point out how our results get
modified on including the saturation.

We first compute the differential rotation and the meridional circulation for $1 M_{\odot}$ stars with rotation periods
of 1, 2, 3, 4, 5, 7, 10, 15, 20, 25.38 (solar value) and 30 days.  Then we run our dynamo code
for all these cases. We
find that our code relaxes to give periodic solutions corresponding to activity cycles in all these
cases.  The butterfly diagrams obtained for the rotation periods of 1, 5, 15 and 30 days are
shown in Figure~6. The contours indicate the values of the toroidal field at the bottom of the
convection zone, whereas the colors indicate the values of the radial field at the stellar surface.
Since starspots are expected to form when the toroidal field at the bottom of the convection
zone is sufficiently strong, the contours can be taken to indicate the butterfly diagrams of
starspots appearing on the surface of the star.  While these butterfly diagrams tend to be
confined to lower latitudes when the rotation is fast, they extend to higher latitudes for slow
rotators like the Sun.  While this does not fit the solar observations in detail, we find that
many of the broad features of solar observations are reproduced.  At lower latitudes in all the
cases shown in Figure~6, we find an equatorward propagation. The toroidal field produced at the
bottom of the convection zone is advected by the equatorward meridional circulation there and
the poloidal field produced from it by the \bl\ mechanism also shows a tendency of equatorward
shift.  On the other hand, we find a poleward propagation in the higher latitudes.  The poloidal
field produced near the surface by the \bl\ mechanism is advected by the meridional circulation
in the poleward direction at higher latitudes. Since diffusion is important in our dynamo model
and the poloidal field diffuses to the bottom of the convection zone where the toroidal field
is produced from it (Jiang et al.\ 2007), we see that the toroidal field also
shows a tendency towards poleward shift at higher latitudes.  This overall tendency of equatorward
propagation in low latitudes and poleward propagation in high latitudes is consistent with
solar observational data. It may be noted that even in detailed solar dynamo calculations
it is nontrivial to confine the butterfly diagram to lower latitudes.  Nandy \& Choudhuri (2002)
pointed out that a meridional circulation penetrating below the bottom of the convection zone
helps in confining sunspot eruptions to low latitudes. Such a penetrating meridional
circulation is used in the majority of solar dynamo papers from our group, but that is not
the case in this paper.  Some recent authors (Mu\~noz-Jaramillo et al.\ 2009; Hotta \& Yokoyama 2010)
used an expression for the $\alpha$-coefficient which is strongly suppressed in high latitudes,
which is not done here.

\begin{figure*}
\includegraphics[width=0.9\textwidth]{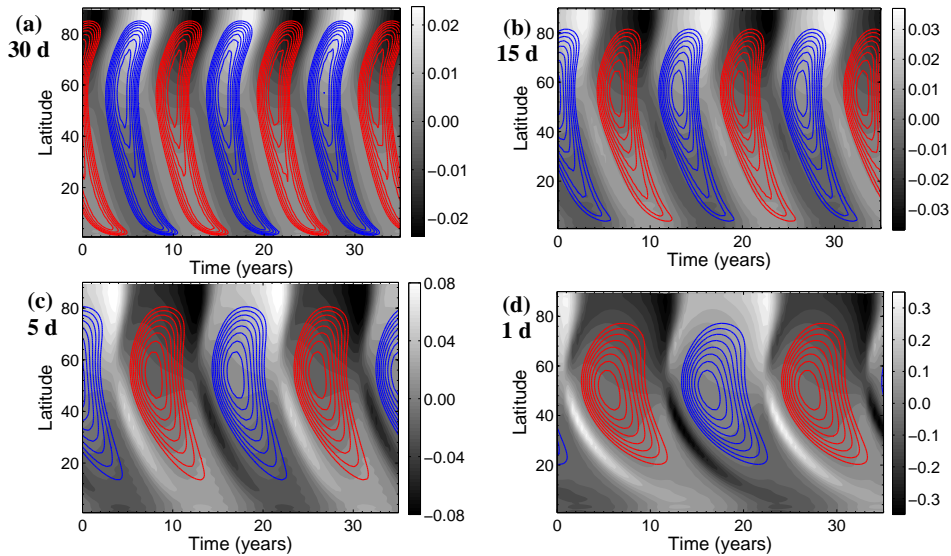}
\caption{Butterfly diagrams of stars with rotation periods 30, 15, 5 and 1~days. Blue (solid) contours correspond to the
positive values of the toroidal field at the bottom of the convection zone, whereas red (dashed) contours correspond to
the negative values of the toroidal field.
The grayscale on the background shows the radial field
(in the unit of $B_0$) on the stellar surface; white denotes the positive value and black denotes the negative.}
\label{bfly}
\end{figure*}

It may be noted that the activity cycle periods for slow rotators like the Sun
are somewhat shorter than the sunspot cycle periods.  The reason behind this would be
obvious on comparing our Figure~3 with Fig.~2 of Chatterjee
et al.\ (2004).  It is clear that the equatorward return flow of the
meridional circulation in the model of KO11 is more
concentrated compared to the meridional circulation we used in our solar dynamo
models (Chatterjee et al.\ 2004; Karak 2010; Karak \& Choudhuri 2011).
As a result, the equatorward flow at the bottom of the convection zone in the model
of KO11 which we use here is stronger than what it was in our
earlier dynamo calculations.  Since the cycle period in a flux transport dynamo decreases
with increasing meridional circulation (Dikpati \& Charbonneau 1999), it is not surprising
that the meridional circulation based on the model of KO11
makes the activity cycle period somewhat shorter than what it is for the Sun. Kitchatinov
\& Olemskoy (2012b) found that the dynamo model based on their meridional circulation gives a
cycle period closer to the sunspot period on including diamagnetic pumping, which is
not included in the present calculations.  In summary, the model developed in this paper,
while applied to the solar case, may not fit all the observational data in quantitative
detail, but the various qualitative features are broadly reproduced.  The main advantage
of the dynamo model presented in this paper is that it gives periodic activity cycles
over a wide range of rotation periods of solar-like stars. We believe that this
model gives a good idea of the general trend in the behavior of the dynamo when the
rotation period is changed.

\begin{figure}[!h]
\centering
\includegraphics[width=0.5\textwidth]{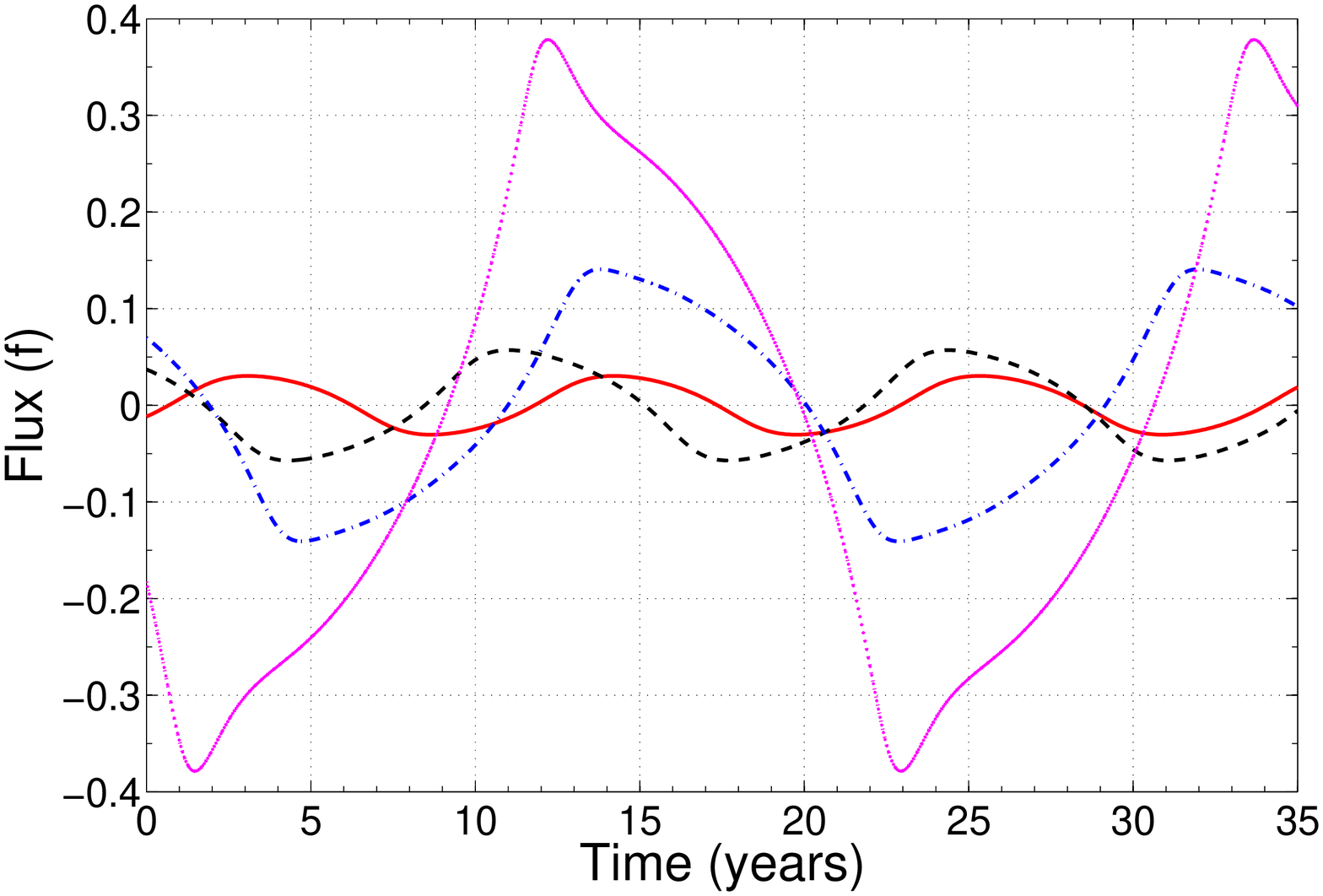}
\caption{Total flux $f$ of the toroidal field calculated over the whole convection zone of stars.
The solid (red), dashed (black), dash-dotted (blue), and dot-pointed (magenta) lines correspond
to stars with rotation periods 30, 15, 5 and 1 days respectively.}
\label{flux}
\end{figure}

\begin{figure}[!h]
\centering
\includegraphics[width=0.55\textwidth]{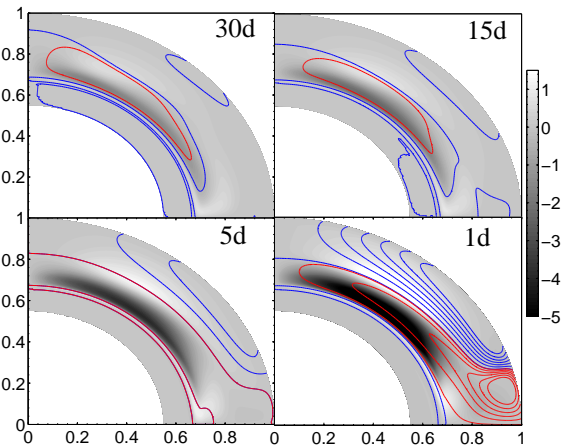}
\caption{Snapshot of the toroidal field (background colors) and the poloidal
field (contours) when the toroidal field in the tachocline
reaches the maximum value. Magnetic fields are in the unit of $B_0$.
The four panels correspond to stars with the four different rotation periods
30, 15, 5 and 1 days. Note that same scales for the toroidal field (color)
and the poloidal field (contours) are used in all the panels.}
\label{torpol}
\end{figure}

For all the rotation periods for which we have carried out dynamo calculations,
we study how the total toroidal flux through the convection zone changes with time.
Writing the total toroidal flux as $f B_0 \Rs^2$, we take $f$ as a measure of the total
toroidal flux. Figure~\ref{flux} shows the variations of $f$ with time for the rotation periods
of 1, 5, 15 and 30 days. We find that the flux varies periodically going through
positive and negative values, as expected from the fact that the toroidal field changes
its direction from one half-cycle to the next half-cycle. It is clear from Figure~\ref{flux} that
the amplitude $f_m$ of the toroidal flux increases with decreasing rotation period
(i.e.\ for faster rotators).  To have an idea of the nature of the toroidal field
generated by the dynamo, Figure~\ref{torpol} gives the distributions of the toroidal field at the
instants when its value in the tachocline is maximum.

We expect more emission in Ca II H/K or in X-ray when there is more magnetic flux.
Since the production of the emission usually involves magnetic reconnection of one flux system
with another, we may naively expect the emission to go as the square of the magnetic
flux, i.e.\ as $f_m^2$.  Now we explore how $f_m^2$ changes with the rotation period
and whether this is consistent with the observational data of Ca II H/K and X-ray emissions
from solar-like stars.  Noyes et al.\ (1984a) discovered that all the data points
for Ca II H/K emission lie in a narrow range if one plots the emission against the
Rossby number rather than the rotation period.  In the present study, we have carried
out calculations for stars with the same mass $1 M_{\odot}$, for which the convective
turnover time will not vary much with the rotation period.  Although in the present
case it would be completely satisfactory to study the variations of $f_m^2$ with the
rotation period, we divide the rotation period by the convective turnover time to
obtain the Rossby number $Ro$.
The convective turnover time was estimated from the local mixing-length relations 
(cf.\ Eq.\ (8) of Kitchatinov \& Olemskoy 2012a).
The turnover time is depth dependent.
The value for the middle of convection zone
($r = 0.86R_\mathrm{s}$) was used to define the Rossby number.
We study the variations of $f_m^2$ with the Rossby
number $Ro$ so that our results can be directly compared with the observational data.
The straight line (dot-dashed) in Figure~9 shows how $f_m^2$ varies with the Rossby number $Ro$.
Since a straight line in this log-log plot is a good fit, we conclude that there is a
power-law relation between $f_m^2$ and the Rossby number $Ro$:
$$ f_m^2 \propto Ro^{-\delta}. \eqno(15)$$
Our simulations give the value $\delta \approx 1.3$, which is the slope of the straight
line in Figure~9.

\begin{figure}[!h]
\includegraphics[width=0.52\textwidth]{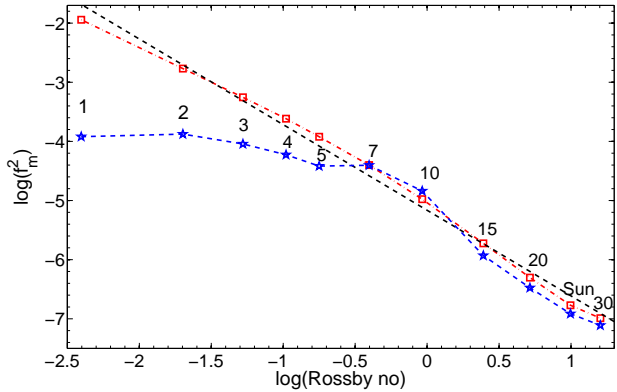}
\caption{Theoretically computed quantity $f_m^2$ as a function of the Rossby number,
$f_m$ being the dimensionless amplitude of the toroidal flux over the whole convection zone.
The values shown by red color (rectangle) and the linear fit (black dashed line)
correspond to the case when $\alpha_0$ is varied according to Equation (13), whereas
the values shown by blue color (star) correspond to the case
when $\alpha_0$ is
varied according to Equation (14).}
\label{sat_lum}
\end{figure}

The Ca II H/K data presented in Fig.~8 of Noyes et al.\ (1984a) or the X-ray data
presented in Fig.~2 of Wright et al.\ (2011) can be fitted with a power law for stars
with long rotation periods.  We can take the luminosity as
$$L \propto Ro^{-\gamma} \eqno(16)$$
for long rotation periods. While the canonical value of $\gamma$ is taken to
be around 2, Wright et al.\ (2011) propose a higher value of 2.70.  From Equations (15) and
(16), we conclude
$$L \propto f_m^{2\gamma/ \delta} \eqno(17)$$
If $\gamma$ is taken to be 2, then we find $L$ to go as a power of about 3 of $f_m$.
On the other hand, if $\gamma$ is 2.70, then $L$ would go as a higher power of $f_m$.
It appears from our analysis that $L$ goes as a higher power of $f_m$ than the
power 2 expected from very naive considerations.  However, we are happy that, in spite
of many uncertainties in our model, we get the general trend.  There are ways of bringing
the theoretical model closer to the observational data.  We shall make some comments
on this in the Conclusion.  In this first exploratory study, we just present
results which follow from the most obvious considerations.

\begin{figure}[!h]
\includegraphics[width=0.5\textwidth]{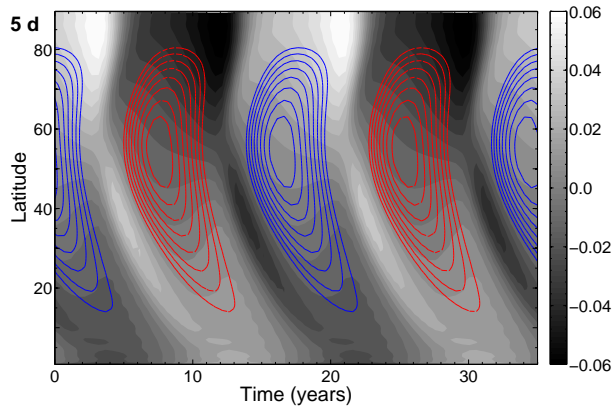}
\caption{Butterfly diagram of a star with rotation period of 5 days. In this case the strength 
of the Babcock-Leighton $\alpha$ is taken according to Equation (14).}
\label{bfly5_sat}
\end{figure}

As we already mentioned, the observational data show a trend of saturation for
stars rotating very fast for which the Rossby number is less than 0.1 (Wright et
al.\ 2011).  In our theoretical model, we do not get any such saturation if the
strength of the \bl\ mechanism is taken to be given by Equation (13).  We now carry on some
calculations using Equation (14) instead of Equation (13). We take $\beta T_s = 3.6257$ days.  Then we would
expect a saturation for stars having rotation periods shorter than 10 days.  We indeed
find that, on using Equation (14) instead of Equation (13), stars with short rotation periods produce
less toroidal flux.  Figure~\ref{bfly5_sat} shows a butterfly diagram for the rotation period 5 days.
This has to be compared with one of the butterfly diagrams in Figure~6.  The dashed
line in Figure~9 shows how $f_m^2$ varies with the Rossby number on using Equation (14) instead
of Equation (13).  It is clearly seen in Figure~9 that on including a saturation in the \bl\ mechanism
there is a tendency of $f_m^2$ growing more slowly and reaching saturation, implying that the
emission also would be saturated for fast rotators in accordance with the observational
data.  We thus conclude that our theoretical model is in qualitative agreement with
the broad features of the observational data.

At last, we come to the question how the activity cycle period varies with the
rotation period. Figure~11 shows how the activity cycle period changes with the rotation
period, the blue open circles giving the results obtained by using Equation (13) and the red solid
circles giving results obtained by using (14).  We find that the cycle period increases
with decreasing rotation period.  It is not difficult to give a physical explanation
of this.
Cycle period in advection-dominated dynamos is largely controlled by the meridional flow. The flow is increasingly concentrated in the boundary layers near the top and bottom of the convection zone as rotation rate increases (Figure~3). The flow velocity in the boundary layers also increases but the layers become thinner so that the flow in the bulk of the convection zone weakens. \citet{M05} argues that angular momentum transport by meridional flow and Reynolds stress balance each other. The Reynolds stress - its viscous part as well as the $\Lambda$-effect - grow weaker than in linear proportion to the rotation rate in the rapid rotation limit due to rotational quenching of the turbulence intensity. \citet{Bea08} found a similar trend with 3D simulations. As a result, meridional flow velocity in the bulk of convection zone decreases.
Since such a meridional circulation is less effective
in advecting the magnetic fields, we find longer cycles for shorter rotation periods.
This theoretical result goes against the observational trend that stars with longer
rotation periods tend to have longer activity cycles (Noyes et al.\ 1984b; Soon \& Baliunas 1994).  We point
out that Jouve et al.\ (2010) also found an increase in cycle period with decreasing
rotation period, exactly like what we have found, contrary to observations. However,
Do Cao \& Brun (2011) have found a solution of this by adding arbitrarily large latitudinal
turbulent pumping in rapidly rotating stars. We believe that some important physics
is still missing from our stellar dynamo models.
In Conclusion we shall discuss some possibilities of closing the gap between
observations and theoretical results.


\begin{figure}[!h]
\includegraphics[width=0.5\textwidth]{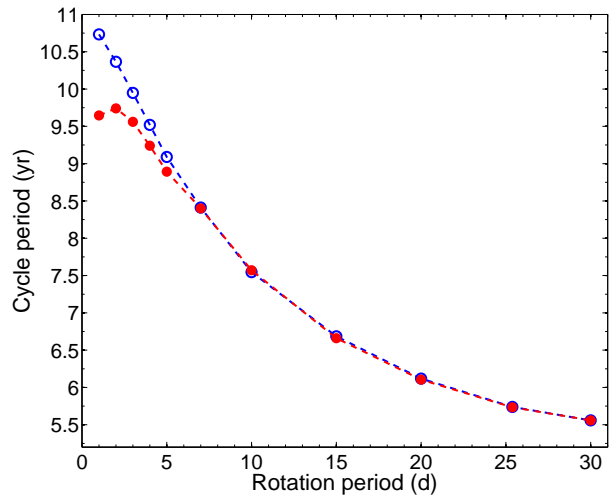}
\caption{Variations of the stellar activity cycle period (in years) with rotation period (in days).
The blue open circles show the case when $\alpha_0$ is varied according to Equation (13), whereas
red filled circles show the case when it is varied according to Equation (14)}
\label{rot_per}
\end{figure}

\section{Results for the rise of magnetic flux parallel to the rotation axis}

The rise of magnetic flux tubes through the solar convection zone has been studied extensively
on the basis of the thin flux tube equation (Spruit 1981; Choudhuri 1990).  When the rotation
period is less than the dynamical time scale of rise due to magnetic buoyancy, the Coriolis
force diverts the flux tubes to rise parallel to the rotation axis (Choudhuri \& Gilman 1987;
Choudhuri 1989; D'Silva \& Choudhuri 1993; Fan et al.\ 1993; Weber et al.\ 2011).  Some mechanisms have been
suggested for suppressing this effect of the Coriolis force, such as the strong interaction
with the surrounding turbulence in the convection zone (Choudhuri \& D'Silva 1990) or the
initiation of the Kelvin--Helmholtz instability inside the flux tube (D'Silva \& Choudhuri 1991).
However, these mechanisms are effective only if the cross-section of the flux tube is rather
small.  Some of the starspots are much larger than sunspots, so these mechanisms are unlikely
to be very important.  Also rapidly rotating stars tend to have polar spots---presumably
diverted there by the action of the Coriolis force (Sch\"ussler \& Solanki 1992).

Both the Ca II H/K emission and X-ray emission from solar-like stars tend to saturate when
the Rossby number is less than about 0.1, as seen in Fig.~8 of Noyes et al.\ (1984a) or Fig.~2
of Wright et al.\ (2011).  This is also approximately the Rossby number below which the rotation
period is shorter than the dynamical time scale. One question which we explore here is whether
the observed saturation could be caused by the effect of the Coriolis force.  In all the
calculations of the previous Section, we took the source function to be given by Equation (11), which
implied that the flux tubes rose radially.  Now, when the rotation period is less than 15 days,
we replace (11) by
 $$S(r, \theta; B) = \frac{\alpha(r,\theta)}{1+(\ov B(r_t,\theta_b)/B_0)^2} \ov B(r_t,\theta_b), \eqno(18)$$
where $\ov B(r_t,\theta_b)$ is the value of the toroidal field
radially averaged over the tachocline from $r = 0.685 \Rs$ to $r=0.715 \Rs$ not at the latitude $\theta$
where the source function is calculated, but at the latitude $\theta_b$ from which a rise parallel
to the rotation axis would bring to flux tube at the latitude $\theta$ when it reaches the stellar
surface.  We obviously have
$$r_b \sin \theta_b = \Rs \sin \theta, \eqno(19)$$
where $r_b$ is the value of $r$ at the bottom of the convection zone.  When we calculate the
source function for stars with rotation periods shorter than 15 days, we now use Equation (18) with $\theta_b$
given by (19). Since $\theta_b$ cannot be larger than $\pi/2$, it is clear from (19) that the
source function vanishes when $\theta$ is larger than $\sin^{-1} (r_b/\Rs)$.  In other
words, the source function $S(r, \theta; B)$ is non-zero only in high latitudes and, in accordance with
Equation (2), the poloidal field generation takes place only in high latitudes.
Therefore by implementing this idea we expect the generation of poloidal field becomes weaker
which may be responsible of producing the saturation of the Ca~II H/K emission. In addition, this
may make the magnetic cycle periods shorter because of restricting the dynamo in shorter domain.

In \S3 we presented dynamo calculations for $1 M_{\odot}$ stars with rotation periods
of 1, 2, 3, 4, 5, 7, 10, 15, 20, 25.38 (solar value) and 30 days.  In the cases of rotation
periods of 15, 20, 25.38 (solar value) and 30 days (slow rotators), the effect of the Coriolis force is not
expected to be very strong.  So the magnetic flux would rise radially and the results of
\S3 would not change even if we take the Coriolis force into account.  Only for stars with
rotation periods of 1, 2, 3, 4, 5, 7 and 10 days (fast rotators), we expect the magnetic
flux to rise parallel to the rotation axis when the effect of the Coriolis force is included.
So we now carry on calculations only for stars with these rotation periods by using the source
function given by Equations (18) and (19) rather than Equation (11).

Figure 12 presents the butterfly diagram for the rotation period of 5 days.  On comparing with
the butterfly diagram for this rotation period based on the assumption of radial rise, as presented
in Figure~6, we find that the magnetic fields are now more confined in the higher latitudes.
This is certainly expected, given that the source function now vanishes at low latitudes.
However the interesting thing is that now the toroidal field is stronger at high latitudes, which
can produce the polar starspots. Otherwise if the toroidal field is not stronger at high latitudes,
then the magnetic flux tubes from the low latitudes go to just parallel to the rotational axis and
not able to produce the polar spots.

We calculate the dimensionless toroidal flux amplitude $f_m$ for all the fast rotators by
treating the source function according to Equations (18) and (19).  Figure~13 shows a plot of $f_m^2$ as
a function of the Rossby number, in which the values of $f_m$ for slow rotators (rotation periods
of 15, 20, 25.38 and 30 days) are the same as used in Figure~9, but for fast rotators
(rotation periods of 1, 2, 3, 4, 5, 7 and 10 days) $f_m$ is calculated by taking the source
function to be given by Equations (18) and (19). For comparison we have overplotted the earlier results of
the radial rise on this plot.  A look at Figure~13 shows that the line
joining the fast rotators gets slightly shifted below when the rise parallel to the rotation
axis due to the Coriolis force is taken into account.
This means that the amount of toroidal flux produced in the fast rotators
is somewhat less when the magnetic flux is assumed to rise parallel to the rotation axis.
This is in agreement with what we expect on the ground that the magnetic fields now occupy
a smaller region (only the high latitudes) of the stellar convection zone.

Although the line for fast rotators is slightly displaced with respect to the line for slow
rotators in Figure~13, we see the same trend of $f_m^2$ increasing with the Rossby number
even for fast rotators that we see for slow rotators. One of our aims was to check whether
the effect of the Coriolis force can explain the saturation of Ca II H/K and X-ray emission
for low Rossby numbers. From Figure~11 we conclude that, although the rise parallel
to the rotation axis due to the Coriolis force causes a decrease in the flux, the general trend of
$f_m^2$ increasing with decreasing Rossby number is not halted by the Coriolis force.  We
presumably need something like the saturation of the Babcock--Leighton mechanism for fast rotation as given
by Equation (10) in order to explain the observed saturation.

\begin{figure}[!h]
\includegraphics[width=0.5\textwidth]{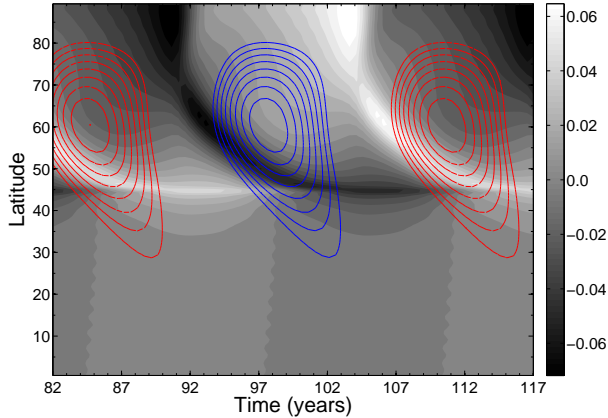}
\caption{Butterfly diagram of a star with rotation period of 5 days. In this case the
source function is given by Equations~(18) and (19) corresponding to magnetic flux rising parallel
to the rotation axis.}
\label{bf_pararise}
\end{figure}

\begin{figure}[!h]
\includegraphics[width=0.45\textwidth]{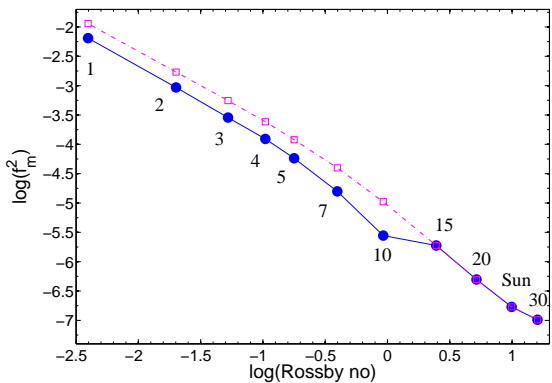}
\caption{The dashed line (magenta points) is the same as in Figure~\ref{sat_lum}, whereas the solid line
(blue points) is for the case where the source function is given by Equations~(18) and (19) for
rapidly rotating stars with rotation periods less than and equal to 10 days.}
\label{diff_buoyancy}
\end{figure}

\section{Conclusion}
Following the success of the flux transport dynamo
model in explaining various aspects of the solar cycle
(Charbonneau 2010; Choudhuri 2011), we explore the
theoretical possibility that similar flux transport dynamos
operate in solar-like stars. We need profiles of
the differential rotation and the meridional circulation
to construct flux transport dynamo models. So we first
compute these flows inside $1M_{\odot}$ stars rotating at different
rates by using the mean-field hydrodynamics model
of KO11. Then we use our dynamo code Surya to construct
dynamo models of these stars. The only nonlinearity
in our model is the standard $\alpha$ quenching
which is sufficient to produce stable solutions
over the parameter regime we have studied. Another
possible source of nonlinearity is the back-reaction of
the dynamo-generated magnetic field on the large-scale
flows such as the differential rotation and the meridional
circulation (see, e.g., K\"uker et al. 1999; Rempel 2006).
In this preliminary investigation of stellar dynamos, we
have not included this back-reaction. This is certainly
justified for slow rotators like the Sun, on the ground
that the amplitude of torsional oscillations (the cyclic
variation in differential rotation) is small compared
to the absolute value of the rotational velocity (Basu \&
Antia 2003; Howe et al.\ 2005; Chakraborty et al.\ 2009)
and the variation of the meridional circulation with the
solar cycle is also not so significant (Chou \& Dai 2001;
Hathaway \& Rightmire 2010; Karak \& Choudhuri 2012).
However, for $1M_{\odot}$ stars rotating much faster,
we have seen that the magnetic field generated is much
stronger and its back-reaction may no longer be negligible.
This effect should be explored in the future.

Our results are in qualitative agreement with many aspects of
observational data.  For example, we find that the dimensionless amplitude $f_m$
of the toroidal flux increases with increasing rotation. We naively expect
the Ca H/K and X-ray emissions to go as $f_m^2$.  We find that we can match
the observational data if we assume the emissions to go as somewhat higher powers
of $f_m$.  However, if $f_m$ were to increase more rapidly with rotation than
what is predicted by our present model, then we see from Equations (15) and (17) that
it may be possible to make the
emission $L$ go as $f_m^2$.  We point out that we have assumed
the Babcock--Leighton mechanism strength to go as $T^{-1}$ as seen in Equation (13).  If we assume this
strength to increase faster with increasing rotation, then we would expect to find the
power law index in Equation (15) steeper.  As seen from Equation (17), this would lead to a weaker
dependence of $L$ on $f_m$. Given the many uncertainties in the theoretical model,
we do not attempt such fine tuning in this paper.  We merely show that the simplest possible theoretical
model qualitatively gives the general trend seen in the observational data.
Allowing magnetic flux to rise parallel to the rotation axis due to the
Coriolis force when the rotation is faster does not change the results
qualitatively.

One disagreement with observational data is that our model predicts that the
magnetic cycles have longer periods when rotation periods are shorter. As we
pointed out, the theoretical model of Jouve et al.\ (2010) also had this difficulty.
We have discussed the reason behind this. The model of KO11
predicts that the meridional circulation is more confined to the
peripheries of the convection zone as the star rotates faster and is less
effective in advecting the magnetic fields.  This less effective
meridional circulation makes the period of the flux transport dynamo longer.
The decrease of cycle period with faster rotation probably implies that the
meridional circulation remains more effective in faster rotating stars than what our
present model suggests.  We have taken the anisotropy parameter $C_{\chi}$ of Equation
(8) to be equal to 1.5 in all our calculations.  This value of  $C_{\chi}$ gives
a good agreement with helioseismology for a  $1 M_{\odot}$ star with solar rotation.
However, it is certainly possible that this anisotropy increases with faster rotation.
A stronger anisotropy may make the meridional circulation more effective in faster
rotating stars and thereby decrease the dynamo cycle period.  We plan to explore
these possibilities in future. At present, even our understanding of the
meridional circulation of the Sun is fairly incomplete.  There have been
recent claims that the meridional circulation of the Sun may have a multi-cell
structure (Zhao et al.\ 2013; Schad et al.\ 2013). Hazra et al.\ (2014) have shown that a flux
transport dynamo can still work with a multi-cell meridional circulation
as long as there is an equatorward flow at the bottom of the convection
zone.  All the calculations in this paper, however, are based on simple
single-cell meridional circulation predicted by the model of KO11 on assuming $C_{\chi}= 1.5$.

By comparing observations with the results of our preliminary theoretical
studies, we find that magnetic effects probably grow somewhat faster with rotation
than what is suggested by our present calculations. Only if the dimensionless
amplitude $f_m$ of the toroidal flux increases faster with rotation, we would
be able to make emission go as $f_m^2$.  Probably flows inside faster rotating
stars also remain stronger than what is suggested in the present model, to ensure
that we have faster dynamos with shorter cycle periods.  We must remember
that in the present dynamo model the generation of the poloidal field from
helical turbulent (the so-called $\alpha$-effect) is not included, since we know from observations that
the Babcock-Leighton process is the major source of the  poloidal field in the Sun.
However, if this is not so true for the rapidly rotating
stars and if the $\alpha$-effect starts contributing to the poloidal field generation in addition
to the Babcock-Leighton process, then that can make the stellar activity stronger
and can also make the cycle periods shorter. Another thing to remember is that
we do not vary the turbulent diffusivity in all the calculations. However, if the
turbulence is weaker in the rapidly rotating stars, possibly due to the rotational or magnetic
quenching (Kitchatinov et al.\ 1994; Karak et al.\ 2014b), then the weaker
turbulent diffusivity can make the dynamo stronger. Therefore, one of the aims of any future
study should be to explore various effects that may make magnetic activity grow
faster with rotation than what we have found. The encouraging thing is that
the present calculations show the qualitative trend of magnetic activity
increasing with rotation.  In what ways the manifestations of stronger magnetic
activity may differ from the manifestations of solar activity is another
important question to be addressed. Some fast-rotating solar-like stars are
known to have starspots much larger than sunspots (Strassmeier 2009).
Some such stars have superflares which are much more energetic than typical
solar flares (Maehara et al.\ 2012).  One related question is whether solar
flares substantially more energetic than the flares recorded so far are
possible in the Sun (Shibata et al.\ 2013).

In this paper, we have restricted ourselves to studying only the regular
aspects of stellar activity cycles. The observational data of stellar
activity presented by Baliunas et al.\ (1995) (also see Baliunas \& Soon 1995) show that many stellar cycles
show strong irregularities.  Constructing theoretical models of irregularities
of the solar cycle has been a major research activity in the field of solar
dynamo theory in recent years.  Studying the irregularities of stellar
cycles presumably will be a fertile research field for the future.  One
intriguing question is whether irregularities of stellar cycles show patterns
similar to solar cycle irregularities.  One important aspect of the solar
cycle irregularities is the Waldmeier effect (Waldmeier 1935) that stronger
cycles tend to have shorter rise times.  The data of Baliunas et al.\ (1995) do
not cover a long enough time interval to conclusively ascertain whether stellar cycles
also show the Waldmeier effect.  However, for a few stars, the time variation
plots of Ca H emission presented by Baliunas et al.\ (1995) cover
several cycles.  If we carefully look at the Baliunas et al.\ (1995) plots of some
stars---notably HD 103095 (in Fig.~1e), HD 149661 and HD 26965 (in Fig.~1f), HD 4628,
HD 201091 and HD 32147 (in Fig.~1g)---we see tantalizing hints that stronger
cycles tend to rise faster, suggesting that the Waldmeier effect is present
in stellar dynamos as well.  Karak \& Choudhuri (2011) have shown how fluctuations
in the meridional circulation can give rise to the Waldmeier effect in a flux
transport dynamo.  The tentative hint of the Waldmeier effect in stellar
cycles certainly suggests that the meridional circulations inside solar-like
stars also probably have large fluctuations.
Apart from fluctuations in the meridional circulation, the other source
of irregularities in the solar cycle is fluctuations in the Babcock--Leighton
process (Choudhuri et al.\ 2007; Choudhuri \& Karak 2009; Olemskoy et al.\ 2013).
The combined fluctuations in the meridional circulation and the Babcock--Leighton
mechanism can explain various aspects of grand minima in solar activity rather
elegantly (Choudhuri \& Karak 2012; Karak \& Choudhuri 2013). The data of
Baliunas et al.\ (1995) show grand minima phases of several stars.  Presumably
the physics behind these stellar grand minima is the same as the physics
behind solar grand minima.  We hope that detailed studies of irregularities
in stellar dynamos will be carried out in the future.

\acknowledgements
We thank an anonymous referee for reading our paper very carefully
and for a long list of suggestions, which improved the presentation.
LLK is thankful to the Russian Foundation for Basic Research
(projects 12-02-92691\_Ind and 13-02-00277).
This work was initiated during ARC's visit to Irkutsk funded by a DST-RFBR Indo-Russian Exchange Program.  A partial support
was provided by the JC Bose Fellowship awarded to ARC by Department of Science and Technology (DST), Government of India.

\end{document}